%% file: main_v2.tex
\documentclass{article}
\usepackage[utf8]{inputenc}
\usepackage[margin=2cm]{geometry}
\usepackage{graphicx}
\usepackage{amsmath}
\usepackage{amssymb}
\usepackage{url}
\usepackage{hyperref}
\usepackage{xcolor}
\usepackage{subcaption}
\usepackage{soul}
\usepackage{lscape}
\usepackage[backend=bibtex,citestyle=authoryear,isbn=false,url=false,eprint=false,sorting=none,uniquename=init]{biblatex}
\title{Urban hierarchy and spatial diffusion over the innovation life cycle}
\author{Eszter Bokányi$^{1,2,*}$, Martin Novák$^{1,3}$, Ákos Jakobi$^{4,5}$, Balázs Lengyel$^{1,2}$}

\date{{\footnotesize%
    $^1$ELKH Centre for Economic and Regional Studies; Agglomeration and Social Networks Lendület Research Group, \\%
    Budapest, H-1097, Hungary\\%
    $^2$Corvinus University of Budapest; Laboratory for Networks, Technology and Innovation; Budapest, H-1093, Hungary\\%
    $^3$ELKH Wigner Research Centre for Physics; Budapest, H-1121, Hungary\\%
    $^4$Eötvös Loránd University; Department for Regional Science; Budapest, H-1117, Hungary\\%
    $^5$ELKH Research Centre for Astronomy and Earth Sciences, Spatial Big Data Lab; Budapest, H-1121, Hungary\\%
    $^*$Corresponding author: \href{mailto:bokanyi.eszter@krtk.hu}{bokanyi.eszter@krtk.hu}\\[2ex]
}}

\begin{document}

\renewcommand{\baselinestretch}{1.5}
\setlength{\parindent}{0em}
\setlength{\parskip}{2em}

\maketitle

\abstract{%
    Successful innovations achieve large geographical coverage by spreading across settlements and distances. For decades, spatial diffusion has been argued to take place along the urban hierarchy such that the innovation first spreads from large to medium cities then later from medium to small cities. Yet, the role of geographical distance, the other major factor of spatial diffusion, was difficult to identify in hierarchical diffusion due to missing data on spreading events. In this paper, we exploit spatial patterns of individual invitations on a social media platform sent from registered users to new users over the entire life cycle of the platform. This enables us to disentangle the role of urban hierarchy and the role of distance by observing the source and target locations of flows over an unprecedented timescale. We demonstrate that hierarchical diffusion greatly overlaps with diffusion to close distances and these factors co-evolve over the life cycle; thus, their joint analysis is necessary. Then, a regression framework is applied to estimate the number of invitations sent between pairs of towns by years in the life cycle with the population sizes of the source and target towns, their combinations, and the distance between them.  We confirm that hierarchical diffusion prevails initially across large towns only but emerges in the full spectrum of settlements in the middle of the life cycle when adoption accelerates. Unlike in previous gravity estimations, we find that after an intensifying role of distance in the middle of the life cycle a surprisingly weak distance effect characterizes the last years of diffusion. Our results stress the dominance of urban hierarchy in spatial diffusion and inform future predictions of innovation adoption at local scales.
}

\clearpage

\section{Introduction}
    % Par 1: innovation life cycle, spatial diffusion theories, findings, and niches
    Fuelled by the recognition that innovation drives economic progress \parencite{schumpeter2003theory, solow1956contribution}, the adoption dynamics of new technologies and products over their life cycle have been analyzed and modeled from the middle of the $20^{\mathrm{th}}$ century \parencite{bass1969new, rogers2010diffusion}. Griliches \parencite{griliches1957hybrid} quickly made geography an important field of this discussion by demonstrating that adoption is faster in the proximity of the original location. Later, Hägerstand \parencite{hagerstrand1968innovation} modeled spatial diffusion with a special focus on cities; he distinguished categories of hierarchical (from a larger to a smaller city) and neighborhood (determined by geographical proximity) processes. This classic work was applied in a wide range of contexts including the diffusion of ideas \parencite{gould_1975}, fashion \parencite{luo_du_he_li_michael_niu_2002}, culture  \parencite{bauernschuster_falck_2014}, or e-shopping \parencite{farag_weltevreden_rietbergen_dijst_oort_2006}. The mixture of hierarchical and neighborhood diffusion mechanisms are thought to resemble routing behavior in social media \parencite{leskovec2014geospatial} such that spreading initially occurs over great distances between large cities but becomes more and more local when it reaches smaller towns. However, the separation of hierarchical versus neighborhood diffusion is still problematic, because large cities are typically surrounded by smaller towns \parencite{christaller1966central}, which demands further analysis with data on spreading events.
    
    % Par 2: diffusion in social networks and spatial limitations
    More recently, adoption dynamics in social networks have received growing attention \parencite{strang1993spatial, valente1996social}. In this stream of literature, diffusion is modeled as a complex contagion process, in which the probability of individuals' adoption increases as network neighbors adopt \parencite{centola2007complex}. This model can predict the online spreading of behavior and the diffusion of online products at the scale of countries well \parencite{centola2010spread, karsai2016local, ugander2012structural}. %Although social networks have well-documented spatial structure \parencite{Liben-Nowell11623, takhteyev_gruzd_wellman_2012}, which this model considers, 
    However, local predictions of the complex contagion framework still suffer from systemic biases \parencite{toole2012modeling}. City size and distance from the origin have remained major confounding factors in modeling spatial adoption \parencite{Lengyel2020} and still limit marketing applications at local scales \parencite{trusov2013improving}.
    
    % Par 3: In this paper + routing + Online social networks and justifying the novelty of data
    In this paper, we aim to better understand the spatial diffusion of innovations by separating the role of urban hierarchy and geographical distance. This is done by analyzing a unique database retrieved from a social media platform called iWiW, the most popular online social network in Hungary in the pre-Facebook era, adopted by 30\% of the total population. We can trace the platform's spatial diffusion through the dynamics of geolocated platform registrations, and the data also allows us to observe each spreading event - that are accepted invitations registered users sent to new users - over the whole life cycle of the product from 2002 to 2012. Thus, we can reconstruct the direction of flows with unprecedented spatial granularity, scale, and time horizon.  

   % Par 4: Overlap and hierarchy
   We demonstrate the problems of hierarchical versus neighborhood diffusion categorization by showing their overlaps. In the middle of the life cycle, when adoption accelerates, the most probable spreading path across towns form long chains signaling that hierarchical diffusion emerges across the full spectrum of towns. In this period, large size differences between source and target towns of spreading events coincide with large shares of invitations sent to very small distances in agglomeration areas where invitations tend to cascade from large cities to neighboring small towns. 
   
   % Par 5: Estimation 
   To disentangle hierarchical and neighborhood diffusion, we estimate the number of invitations sent between pairs of towns by years with zero-inflated negative binomial regressions, in which the population sizes of the source and the target towns, their combination, and the distance between them are the explanatory variables. This framework allows us to separate the single effect of city size and distance and also consider their joint effect. By considering the direction of flows, we discover that hierarchical diffusion prevails initially across large towns only but emerges in the full spectrum of settlements in the middle of the life cycle. Interestingly, we discover a characteristic size difference as a sign of hierarchical diffusion at the end of the life cycle. Fixing the urban hierarchy variables at given levels, we find that the role of distance is minor in the beginning, intensifies in the middle, and shades away towards the end of the life cycle. This finding contradicts previous gravity estimations that have suggested monotonous intensification of distance effect.
    
    % Par 6: 
    Our results confirm that spatial diffusion of innovations occurs along the hierarchical order of cities. We provide new insights into the dynamics of the exact mechanism over the innovation life cycle. This new evidence is important for local adoption predictions that aim to overcome the systemic biases of settlement size and geographical distance in network diffusion models.

\section{Results}

    \subsection{Data}
    Our analysis is based on a dataset collected from iWiW, a Hungarian online social networking site launched in 2002. The service was a remarkable innovation in the Hungarian market of online services of its time and became highly popular in 2005. Almost 3 million profiles were registered until 2008 (30\% of total population) when iWiW was the most visited Hungarian webpage. The life cycle of iWiW is depicted by monthly registrations in Supporting Information~1, in which registrations are broken down to age groups as well. 
    The website lost its primacy from 2010, when Facebook overtook the market and the iWiW owner stopped the service in 2014. We have access to the anonymized version of all public profile data on iWiW, including friendship ties, demographics, location, registration date, and date of last login. Data was collected and delivered by the data owner in January 2013. Previous research on iWiW has looked into the spatial pattern of the iWiW social network and its relation with socio-economic phenomena \parencite{lengyel2015geographies, wachs2019social, toth2021inequality}, churn from the website \parencite{torok2017cascading, lorincz2017collapse}, and diffusion \parencite{Lengyel2020, katona2011network}.

    In this paper, we analyze invitation patterns that capture directed and geolocated spreading events at unprecedented scale and time horizon. Until 2012, users could only register on the website if they have been invited by another iWiW user. The individual-level data used in this paper contains the ID of iWiW users, the ID of those users who sent the invitation, self-reported location information of users at the settlement level, and day of registration. Thus, we are able to construct a network of invitations from our dataset, where nodes are the iWiW users, and directed edges between them represent the invitor-invited relationships at the individual level. This way, we create a network with 3,059,363 users and 2,636,779 invitations between them. We then aggregate our individual-level network into the 2555 Hungarian settlements given in the users town of residence field in each year of the life cycle. In this dynamic settlement-level spatial network, we have directed edges with an edge weight corresponding to the number of invitations going from the source settlement to the target settlement in the given year. We omit self-loops from the analysis, where the settlement of the source node and the target node are identical. 

    \subsection{Overlap of hierarchical and neighborhood diffusion}
    
    \begin{figure}[b!]
    \centering
    \includegraphics[width=\linewidth]{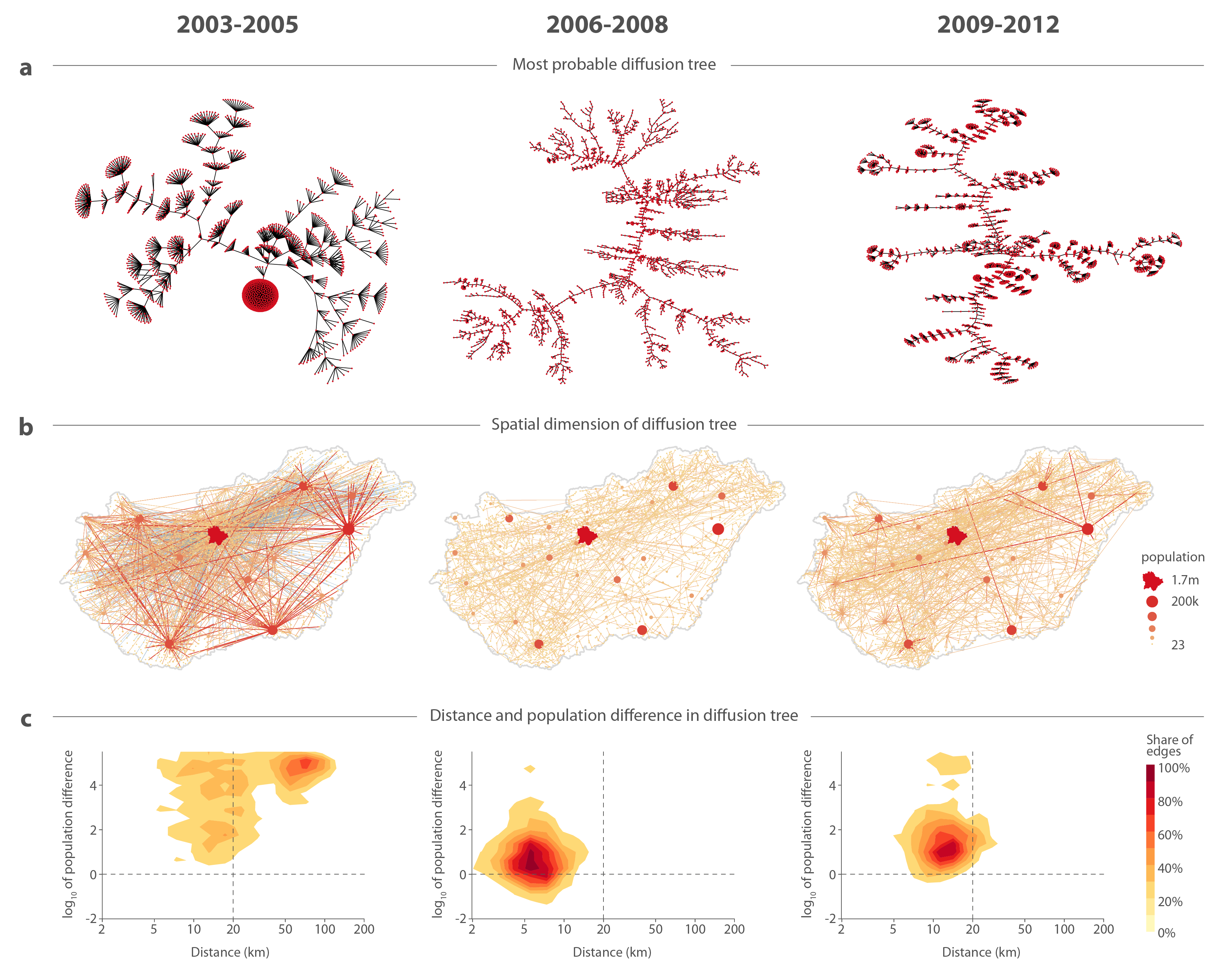}
    \caption{\textbf{Hierarchical and neighborhood diffusion over the innovation life cycle.} \textbf{(a)} Most probable invitation paths given by the solution of the minimum weight branching problem on the transformed network with $\tilde{p}_{ST}$ weights. Nodes correspond to settlements.  \textbf{(b)} Minimum weight invitation trees of the first row with settlements positioned on the map of Hungary. Edges are colored according to the size of the source settlement (edges with Budapest as a source are blue), that is also indicated by the size of the nodes. \textbf{(c)} Distribution of tree edges with respect to two measures: distance between source and target settlements (horizontal axis), and $\log_{10}$ of source and target population size fraction (vertical axis). The vertical black line separates tree edges with less than 20~km distance between source and target, the horizontal black line separates tree edges that go down the settlement hierarchy (source size is larger than the target size), and that go up the settlement hierarchy (target size is larger than the source size). The top left quadrant corresponds to edges that are both sent to a very close distance and in a downwards hierarchical pattern.}
    \label{fig:fig1}
    \end{figure}    
    
    To demonstrate the problem that hierarchical and neighborhood diffusion often overlap, Figure \ref{fig:fig1} illustrates the most probable invitation tree in the spatial diffusion measured from the settlement-level network across the iWiW life cycle. We denote the number of invitations in a given year going from settlement $S$ (source) to settlement $T$ (target) by $w_{ST}$. To identify most probable diffusion paths, we normalize $w_{ST}$ by the number of total outgoing invitations such that edges represent outgoing invitation probabilities $p_{ST}$, where $ p_{ST} = \frac{w_{ST}}{\sum_T w_{ST} }$. The probability of a cross-settlement diffusion path is the product of these $p_{ST}$ values along the path edges. We normalize weights by taking the logarithm and multiply by -1 so that the new weights are $\tilde{p}_{ST} = -\log \frac{w_{ST}}{\sum_T w_{ST} }$. This transformation makes the maximum probability tree of the graph the one for which the sum of the weights is minimal. Using the modified weights enables us to search for the most probable invitation tree \parencite{PastoreYPiontti2014, schlosser2020covid} by detecting the minimum weight branching in the invitation network \parencite{tarjan1977finding}. We use the efficient implementation of the Chu-Liu-Edmonds algorithm for obtaining the minimum weight spanning tree by Gabow \parencite{Gabow1986}. 

    Figure \ref{fig:fig1}a reports a characteristic change in the pattern of major flows of diffusion across settlements. In 2003-2005, the paths in the tree are short such that leaves can be reached through only a low number of internal vertices. The tree even includes a node that has an especially large degree corresponding to Budapest that acts as a major source of invitations during this period. These observations suggest fast diffusion cascades, in which few settlements stand out in distributing the innovation. However, the pattern changes in 2006-2008 when paths in the diffusion tree form long chains, and large-degree nodes disappear signaling a smoother and more equal diffusion through the whole size-spectrum of settlements. This tree with long paths remains in years 2009-2012 as well but smaller cascades emerge again at the tip of the branches.
       
    The spatial pattern of these diffusion trees illustrated in Figure \ref{fig:fig1}b, provides support to the hierarchical diffusion of Hägerstrand \parencite{hagerstrand1968innovation}. Hierarchical diffusion of iWiW initially happened only from the largest cities that have mediated the innovation to all other settlements even to large distances. Later, from the middle of the life cycle, the long diffusion paths have emerged along the urban hierarchy such that flows of spreading went from slightly larger to slightly smaller settlements to a smaller geographical distance. Finally, at the end of the life cycle, both mechanisms contribute to the diffusion.

    Next, we further demonstrate the co-evolution of the hierarchical and neighborhood diffusion over the life cycle in Figure \ref{fig:fig1}c.  This is done by plotting the geographical distance between settlements connected by the diffusion tree in Figure \ref{fig:fig1}a against the population difference (measured as the $\log_{10}$ of the fraction of the source and the target population sizes). The first period is characterized by long-distance cascades of hierarchical diffusion. This changes to local diffusion across similar settlements in the second period. Finally, the emerging small cascades remain local. 

    Taken together, in the above descriptive analysis we find by looking at the most probable diffusion tree across settlements that hierarchical diffusion and neighborhood diffusion are closely related and co-evolve over the life cycle. Thus, their joint systematic analysis is necessary. Supplementary Information 2 further describes the dynamics of hierarchical and neighborhood invitations.
    
\subsection{Disentangling hierarchical and neighborhood diffusion}
    
According the observations presented above, the effect of urban hierarchy in spatial diffusion must be disentangled from other major geographical factors. The first of such geographical factors is the effect of the urban concentration captured by urban scaling law and the second is the effect of distance captured by the gravity law, both of which have been previously found to characterize spatial diffusion of iWiW \parencite{Lengyel2020}. The scaling effect \parencite{bettencourt2007growth} manifests in disproportionally more invitations sent from large settlements early on in the life cycle compared to what their population size would linearly suggest, but this superlinear scaling disappears throughout the majority and the laggard phases \parencite{Lengyel2020}. The gravity effect captures the intensity of diffusion between settlements $S$ and $T$ as a function of populations $P_S$ and $P_T$ and the term $d^{-\chi}$, where $d$ is the Euclidean distance between the two settlements, and $\chi$ is the gravity coefficient, which is found to have an increasing power on diffusion over the life cycle \parencite{Lengyel2020}.  

We apply a regression framework to capture the role of hierarchical diffusion, the effect that an innovation is more likely to go from larger settlements towards smaller settlements, assuming the co-presence of urban scaling of adoption and the gravity law in sending invitations. In every year, we observe the number of invitations going between settlement pairs $w_{ST}$, but there are even more settlement pairs between which there is no invitation process at all. To account for these excess zero observations, we use a zero-inflated negative binomial regression model \parencite{Long1997}. In this regression technique, we assume that the excess zero counts come from a logit model and the remaining counts come from a negative binomial model specified in Materials and Methods. We can predict invitation counts by including all possible settlement pairs into the data. 

We estimate the following equation for the invitation counts $w_{ST}$:
    \begin{equation}
        \log w_{ST} \approx \alpha (\log P_S)^2 + \beta \cdot (\log P_S \cdot \log P_T) + \gamma (\log P_T)^2 + \delta \log P_S +\varepsilon \log P_T + \chi \log d_{ST} + C,\label{eq:main}
    \end{equation}
where the coefficients $\alpha$, $\beta$, $\gamma$, $\delta$, $\varepsilon$, and $\chi$, and the constant $C$ are given by the maximum likelihood estimator of the ZINB model in Stata \parencite{stata}. To account for the gravity effect, we include the distance term $\log d$, and the population terms corresponding to the source and target populations $\log P_S$ and $\log P_T$. However, Figure~1 suggests that the role of hierarchy changes in the life cycle, because the most probable tree of invitations shifts from larger towns to long hierarchy chains, then to smaller towns dominating the invitations. Therefore, we add second-order terms $(\log P_S)^2$, $(\log P_T)^2$, and $\log P_S\cdot \log P_T$ to the estimation, which can capture and test multiple hierarchical invitation scenarios by using terms that are able to describe a general second-order surface in the variables $\log P_S$ and $\log P_T$. For the full regression tables and the logit part of the model, see regression coefficients and parameters in the tables of Supplementary Information~4. 

We start with the effect of settlement size. We would like to understand how the source and target settlement populations influence invitation counts if we keep the distance term constant. In other words, given two settlement pairs for which the distance is equal, how does the invitation count differ for different $P_S$ and $P_T$ pairs. Because of the second-order terms in the regression, the surface defined by the population terms
    \[ \alpha (\log P_S)^2 + \beta \cdot (\log P_S \cdot \log P_T) + \gamma (\log P_T)^2 + \delta \log P_S +\varepsilon \log P_T +  C\]
is a general second-order surface. For us, the important range from this surface that is defined by the regression coefficients is where both $P_S$ and $P_T$ are within realistic population sizes, that is, if they are larger than a small village ($\approx200$ people), but smaller than the capital city Budapest ($\approx2\cdot 10^6$ people).

    \begin{figure}[t!]
     \centering
     \includegraphics[width=0.8\linewidth]{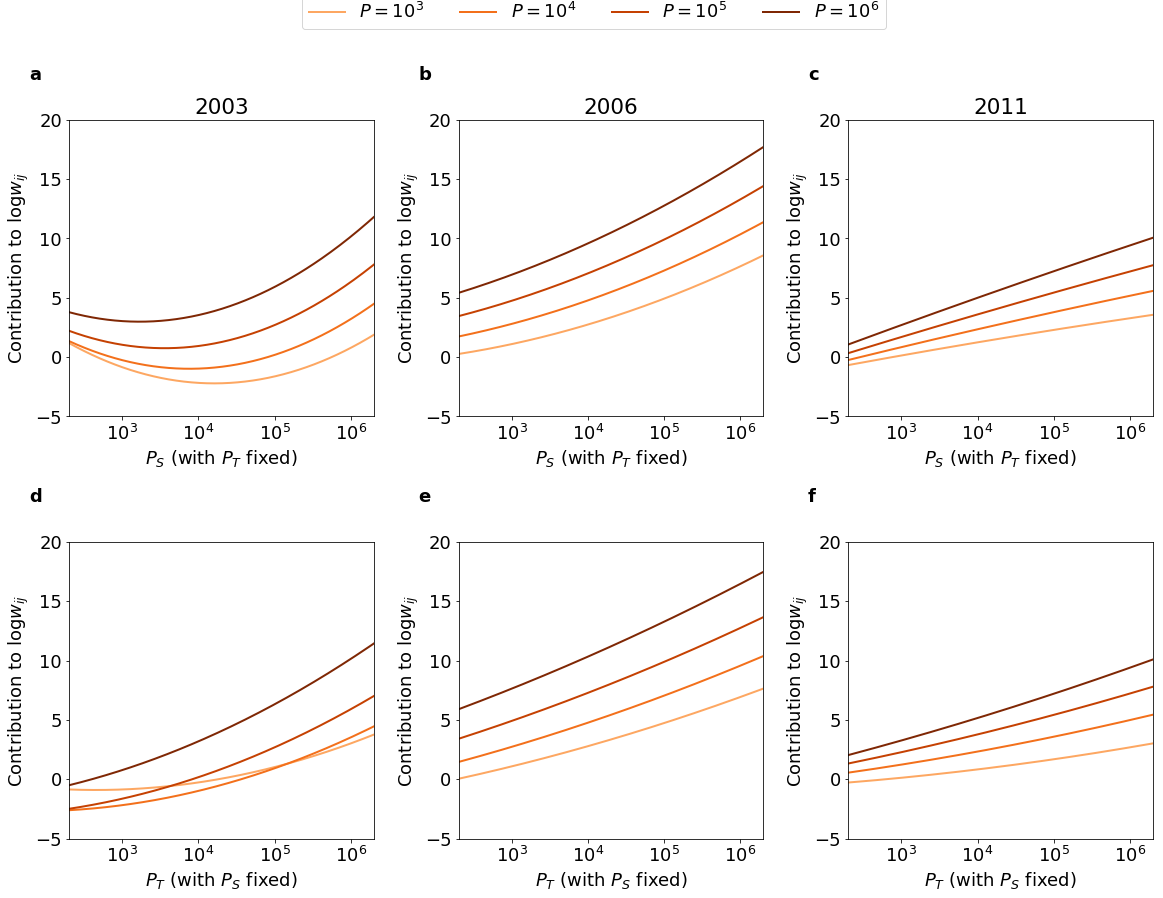}
     \caption{\textbf{Settlement size in spatial diffusion.} The effect of source (a-c) and target (d-f) settlement size with all other variables fixed in the three selected years 2003, 2006, and 2011.}
     \label{fig:size}
    \end{figure}

First, we show how $\log w_{ST}$ changes with $P_S$ for four different target population sizes $P_T$ (Figure~\ref{fig:size}a-c) in three selected years (2003, 2006, and 2011) that are in three different phases of the life cycle. These four different target populations roughly represent a village ($10^3$), a small town ($10^4$), a larger town ($10^5$), and the magnitude of the capital city, Budapest ($10^6$). In 2003, there is a minimum in the curves of all size regimes, meaning that mid-sized source settlements act less as invitors in the network than small villages and large towns. The contribution to the $\log$ invitation count is highest for large towns and the capital, and the relationship is significantly nonlinear between $\log w_{ST}$ and $\log P_s$ (as confirmed by the coefficients of second-order terms in SI3). This suggests that in the beginning of the life cycle of a product, the effect of source settlement population size is even larger than what an urban scaling law would predict. In 2006, $\log w_{ST}$ increases with $\log P_S$ in all four selected $P_T$ regimes, therefore, the larger the source settlement, the more the invitation count for a given target size. In 2011, this monotonicity still holds, although the relationship becomes concave suggesting a faster-than-power-law decrease in the role of large settlements in the diffusion of invitations. For given source settlement sizes (again $10^3$, $10^4$, $10^5$, and $10^6$), panels d-f show that an increasing target settlement size increases invitation counts. The curves from a-f (as well as coefficients in SI4) show that second-order terms are indeed important in estimating the number of invitations, therefore, the observed relationships are rarely linear between $\log w_{ST}$ and $\log P_S$ or $\log P_T$. This means that power-laws between $\log w_{ST}$ and any combination of $P_S$ and $P_T$ are not feasible models for this system.

\begin{figure}[t!]
 \begin{center}
          \includegraphics[width=0.8\linewidth]{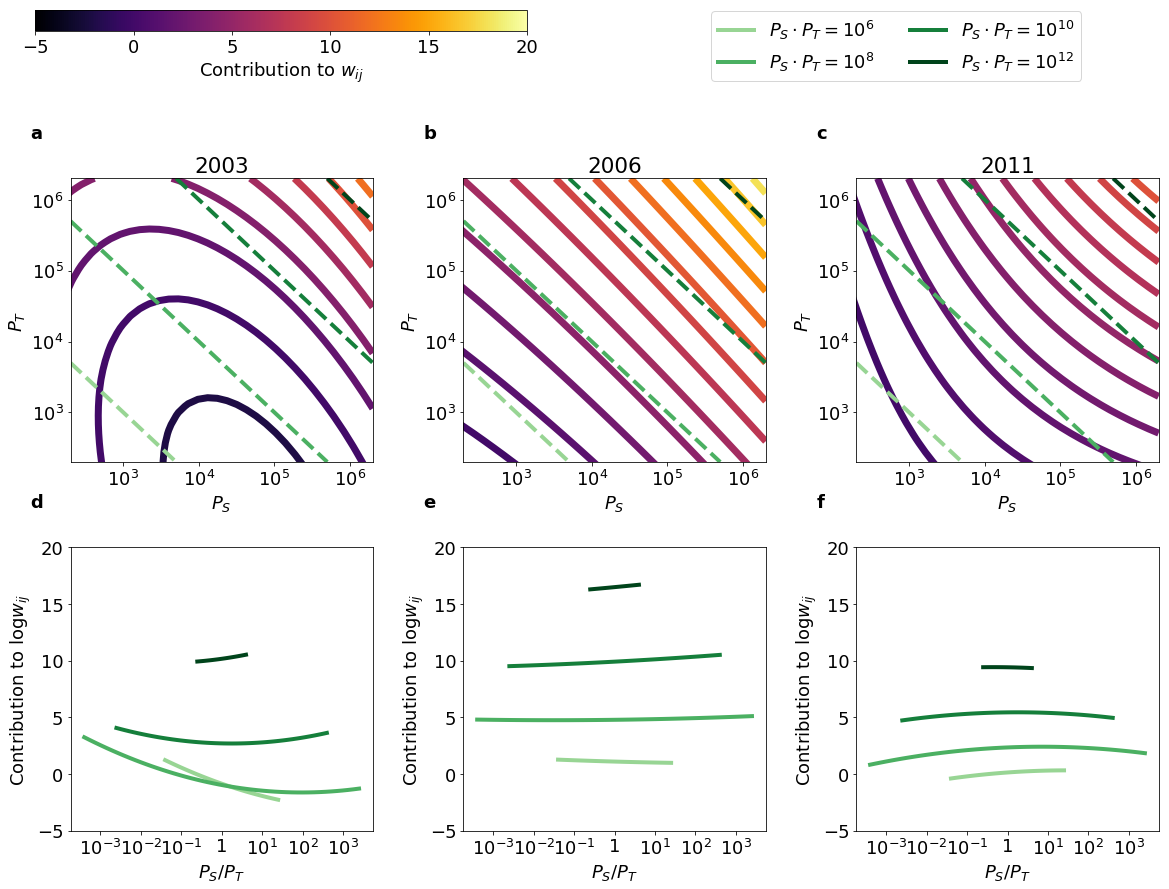}
 \end{center}
 \caption{\textbf{Urban hierarchy in spatial diffusion.} Contour lines in panels (a-c) show level curves of the contribution of the source ($P_S$) and target ($P_T$) settlement sizes to the negative binomial term of the regression with the distance $d$ fixed. Lighter color indicates more invitations, see the colorbar of the top left corner. Green dashed lines correspond to $(P_S,P_T)$ pairs where the number of potential total connections between two settlements is constant, that is, where $P_S\cdot P_T$ is constant, with darker color referring to more potential total connections ($10^4$, $10^6$, $10^8$, or $10^10$). Panels (d-f) show the surface height corresponding to the position of the green dashed lines from panels (a-c). These are the estimated invitation count contributions given that the settlement pair has the same number of potential total contacts and fixed distance.}
 \label{fig:hierarchy}
\end{figure}

But Figure~\ref{fig:size} does not tell us about the effect of hierarchy, since we cannot compare the combined effect and relative position of $P_S$ and $P_T$ on $\log w_{ST}$. Panels a-c on Figure~\ref{fig:hierarchy} present the level curves of the contribution to $\log w_{ST}$ from the population terms and the constant $C$ from equation \eqref{eq:main} in the variable space of $P_S$ and $P_T$. Coloring of the level curves goes from the darkest minimum to the lightest maximum. In the bottom left corner of panels a-c, $P_S$ and $P_T$ both correspond to small village population sizes, whereas the top right corner corresponds to the case where both $P_S$ and $P_T$ are as large as Budapest, the capital city. In between, this area represents every possible (even hypothetical) settlement pairs in Hungary in terms of $P_S$ and $P_T$. As expected, the number of invitations is highest in all three selected years in the top right corner: if both the source and the target settlements are very big. But the finer structure of this estimation is remarkable, and is able to tell us more about the hierarchy effects in detail.    
    
As a most important marker of hierarchy effects, we investigate four different cross-sections of this surface estimated by ZINB that are marked by the green dashed lines of tangent -1 in the upper panels. These green dashed lines correspond to $P_S\cdot P_T = \mathrm{const.}$ pairs, in other words, these lines mark all settlement pairs for which the potential number of all connections is the same. Along this line in panels a-c, in the very middle, there is the case when $P_S=P_T$, but towards the right bottom corner, $P_S>P_T$, and the population share of the source settlement grows. Towards the top left corner, $P_T>P_S$, and the population share of the target settlement grows for the fixed number of total connections. The bottom panels d-f show the surface height along these selected lines, whose shape can be classified into four different categories.
    
\begin{enumerate}
 \item If it is increasing towards large $P_S/P_T$ values, it means that the invitation contribution is larger for those pairs, in which the source settlement is larger than the target, even if there are the same number of total possible connections for other pairs. This corresponds to the hierarchical invitation pattern in which larger settlements tend to invite smaller settlements.
 \item On the other hand, if this curve is monotonically decreasing, then it is exactly the other way round: relatively larger targets and smaller sources lead to more invitations. This corresponds to a reverse effect in which invitations flow back from smaller settlements to larger ones. 
 \item If there is a minimum in this curve within the given realistic size regimes, it means that larger size differences between $P_S/P_T$ are favored in both directions compared to settlements of roughly the same sizes, or that large size differences in either direction lead to more invitations.
 \item A maximum would select one ``favored'' size difference for which invitation count contribution is maximal. 
\end{enumerate}

We can see that the largest size regime always falls into the first category of the above four, here, a relatively larger source settlement in the pair yields more invitations (Case 1). However, in 2003, the two smallest categories, $P_S\cdot P_T=10^6$, and $P_S\cdot P_T=10^8$ both show a strong inclination for reverse hierarchical invitations (Case 2), where the smaller $P_S/P_T$ is, the larger the invitation count. And the regime of cities with roughly 100,000 population disfavors similar sizes: contribution is largest at both ends of the possible $P_S/P_T$ regime (Case 4). In 2006, these remarkably diverging behaviours come close to each other, and in the two largest size regimes, there is a small hierarchical effect (Case 1), the third line has a very shallow minimum, meaning that hierarchy plays almost no role here, and there is a very slight decreasing effect for the smallest settlement pairs (Case 2). In 2011, the three largest curves show a favor for similar settlement sizes (Case 3), and a slight hierarchical effect for the smaller settlement pairs (Case 1).

    \begin{figure}[t!]
     \centering
     \includegraphics[width=10cm]{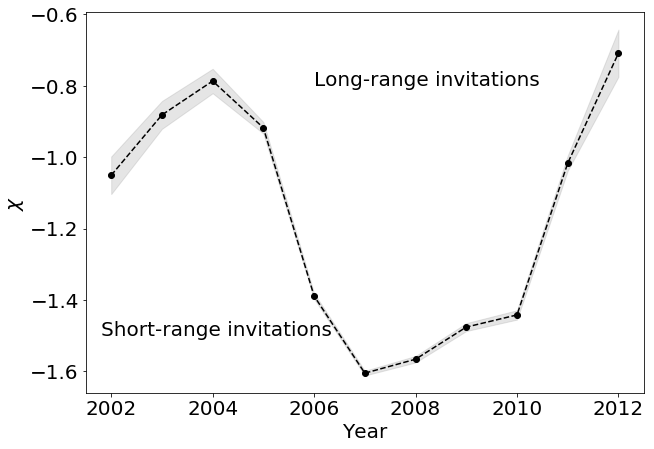}
     \caption{\textbf{Distance effect in spatial diffusion.} Value of the coefficient $\chi$ estimated from the negative binomial part of the ZINB model characterizing the dependence of invitations on spatial distance between two settlements.}
     \label{distance}
    \end{figure}
    
Finally, Figure~\ref{distance} shows the coefficient $\chi$ of the distance term $\log d_{ST}$ from equation \eqref{eq:main} with its estimated error. There is a clear distinction between years with more long-range invitations (2002-2005, 2011-2012), and the short-range invitation years (2006-2010). This result from the regression is thus in line with the intuitive picture from Figure~\ref{fig:fig1}, in which the share of invitations sent to less than 20~km was very high in Figure~1c for most the most probable invitation tree edges in years 2006-2009. Here we can see that this effect remains if we control for excessive zero counts and population sizes.

\section{Discussion}

% PAR 1: findings and novelty
This paper contributes to the traditional discussion on spatial diffusion of innovations by disentangling the role of urban hierarchy from geographical distance using data from unprecedented detail and timescale. Looking at the most probable diffusion paths using direction and timestamp of nearly 3 million spreading events, we discover that diffusion initially cascades from large cities to all other settlements. Later, in the middle of the life cycle when adoption accelerates, hierarchical diffusion manifests in long chains and emerges along the entire spectrum of urban hierarchy. Unlike in previous research that investigated distance effect separately, our multivariate approach unfolds the real effect of geographical distance that intensifies stronger around the adoption peak but becomes weak again in the late phases of the life cycle.

% PAR 2: implementation and diffusion modeling
% implement the emergence: size effect or social network effect? Early Adopter: university students. Peak: local connections and circles. Laggards: stars at the bottom of hierarchy as observed in scaling exponent decreases as well
% implement distance effect

Our results shed new light on the main puzzles of spatial diffusion research: how is the spatial distribution of individuals open to novelty and how does the structure of social interaction networks influence diffusion prediction at local scales? Previous models have found that settlement size and distance are remaining sources of error in complex contagion models. The finding that the role of urban hierarchy is dynamically changing over the life cycle and has a consequence on the role of distance as well might be important for future models. Simulated complex contagion models on social networks should incorporate the hierarchical diffusion processes.% at the settlement level as well as distance dependence at the same time.

% PAR 3: extension to other observations

% discuss different data sources: Twitter registration, Twitter hashtags-retweets, patents-publications
% Sina Aral Science fake news

The consequences of these findings might reach far concerning the potentials of policy interventions in supporting or blocking diffusion. Adoption of new technologies or scientific knowledge is the engine of progress; thus, a policy targeting adoption according to the place of the settlement in the urban hierarchy can benefit the society. However, the costs and motivations of adoption vary greatly across products, and technologies; hence, the generality of spatial patterns reported here must be looked at across various innovation flows. The diffusion of social media is a specific example in which adoption is greatly motivated by social interactions on the website. To prove the generality of the illustrated spatial diffusion patterns, one needs to investigate products that are not strongly associated with the social dimension. Policy might want to block the diffusion of harmful novelty, such as fake news. However, these typically spread very quickly; thus, future research shall compare spatial diffusion of innovations of long and short life cycles. Finally, other socio-economic factors such as development, instrastructure, education, institutions among others might have a role in diffusion that we did not consider in this paper.

\section{Materials and methods}

    \subsection{Data}
Although free registration became possible in 2012, this previously singular option remained the major means of joining the website, with  more than 60\% of iWiW users registering through an invitor even after this date. Thus, for the majority of the users, we have the anonymized identifier of the inviting user as well. Since our data was collected in 2013, the ID of the invitor was missing in those cases when the profile of invitation sender was already deleted or the profile was registered after June 2012 via the free registration option.

Users had to choose a town of residence to be displayed along their profiles. Users could change this location field easily, moreover, the reliability of this piece of information cannot be checked, although the menu offered a limited number of valid location strings for users. Nevertheless, we use this town of residence as geolocation information about both the users and their invitors. 

In our network, nodes are geolocated at the settlement level using the self-reported location field, we leave users having more that one, but less than 5000 connections, and we omit users for whom the first and last login happened on the very same day. Moreover, we only include users with a Hungarian location, since a certain number of the registrations came from neighboring or other European countries. 

Individual data access is restricted  by an NDA between the research group and the data owner. Aggregated data can be requested in email sent to lengyel.balazs@krtk.hu.

\subsection{Most probable invitation paths}

We construct the most probable spreading of the invitations on the settlement networks by looking for the most probable invitation tree using an efficient implementation of the Chu-Liu-Edmonds algorithm \parencite{Gabow1986}. We suppose that the probability of an invitation path is the product of the probabilities along the constituting directed edges, where we normalize invitation counts by the number of total invitations going out of the source settlement in three time periods (2003-2005, 2006-2009, 2010-2012). By taking the logarithm and multipying edge weights by -1, this problem is equivalent to finding the minimum weight branching in the directed weighted invitation network. Because the network is not always strongly connected, we add an artificial source node that we connect to every other node with a sufficiently large edge weight such that the algorithm avoids these edges if possible, but we remove the artificial root node at the end of the process.

\subsection{Regression framework}

The zero-inflated negative binomial regression model is formalized by:

    \begin{align}
    w_{ST} &= e^{k}\cdot\underbrace{\left(1-\frac{e^l}{1+e^l}\right)}_{\mathrm{nonzero~inv.~probability}},\label{eq:fullmodel}\\
    \log w_{ST} &= k + \underbrace{\log \left(1-\frac{e^l}{1+e^l}\right)}_{\mathrm{small}}.\label{eq:contr}
    \end{align}
    
In the first of the above equations (\ref{eq:fullmodel}), $l$ is the log-odds term from the logit model that is used to estimate the propability that there is a nonzero number of invitations between settlement pairs. The parameter $c$ characterizes the actual invitation counts between settlement pairs given that the count is nonzero. The second equation (\ref{eq:contr}) shows that the contribution to $\log w_{ST}$ is mostly contained in the term $k$, since the probability of the invitation between the two settlements being nonzero is close to one, for which the logarithm is close to 0. Both $k$ and $l$ are governed by the processes mentioned before, therefore, we choose the terms that predict them in the ZINB model to reflect these. 

We estimate the terms $k$ and $l$ from Equation~\ref{eq:fullmodel} for each year between 2002-2012 using Stata \cite{stata}.

\section{Acknowledgements}
The suggestions of Ferenc Bencs in identifying most probable invitation paths is gratefully acknowledged. Figure 1 has been finalized by graphic designer Szabolcs Tóth-Zs. The authors acknowledge financial help received from National Research, Development and Innovation Office of Hungary (KH 130502).

\printbibliography

\clearpage

\section*{Supplementary Information}

\section*{Supplementary Information 1: The iWiW life cycle}

\begin{figure}[h!]
        \centering
        \includegraphics[width=0.7\textwidth]{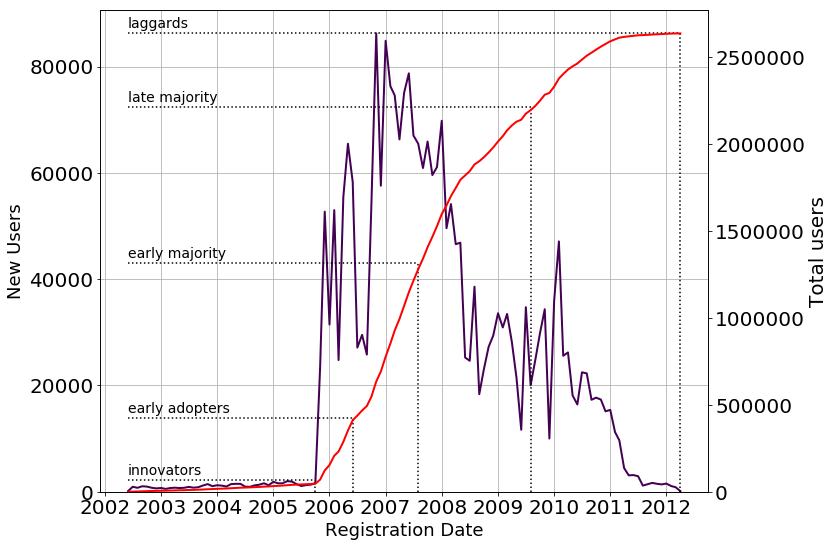}
        \caption{Number of registrations on the iWiW social network by month. Red curve shows the cumulative number of adopters, annotations reflect the adopter categories of \cite{rogers2010diffusion}.}
        \label{fig:sec2_invfreq}
    \end{figure}

The number of registrations was low in the first few years but iWiW reached a several hundred thousands of people by 2005. The largest Hungarian internet provider company acquired the website in 2006, and the number of registered users grew rapidly from 1.5 to more than 4 millions until December 2008. Moreover, between 2005 and 2010, iWiW was the most frequently visited Hungarian webpage. This is well reflected in Figure~\ref{fig:sec2_invfreq}, that shows the number of registrations over the full life cycle of the website, and confirms that at the peak of iWiW's popularity, the number of registered users jumped to more than 50.000 monthly, and increased further with high variability to a peak about 90.000 invitations per month until the middle of 2007.
    
    \clearpage
    
   \noindent  The figure shows how early adoption was driven by the age group of 19-29 years, then people of age 8-18 and 30-40 followed. Older age groups between 41-51, 52-62, then 63-73 gradually joined in the adoption process. This means that age structure and age-related social and family contacts of settlements might influence adoption patterns. 

    \begin{figure}[h!]
        \centering
        \includegraphics[width=0.7\textwidth]{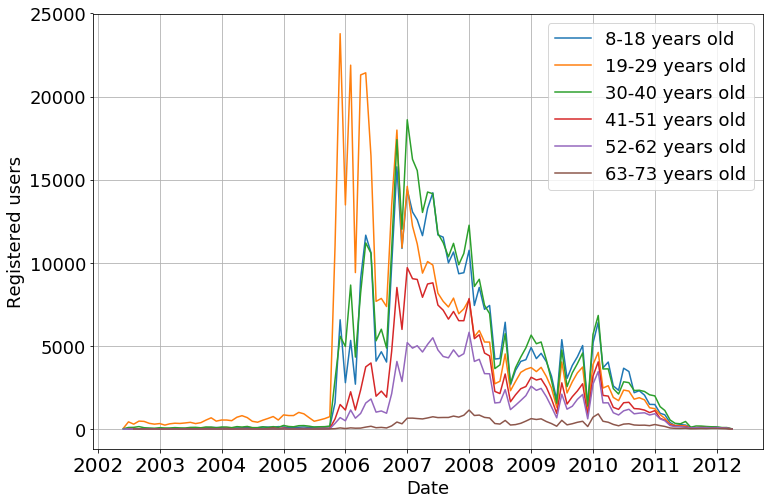}
        \caption{Number of registrations on the iWiW social network by month per age group.}
        \label{fig:agereg}
    \end{figure}
    
\clearpage

\section*{Supplementary Information 2: The share of invitations by diffusion categories}

\noindent The average distance of invitations declines somewhat in 2006-2007, although the standard deviation of invitation distances marked by the errorbars remains high. However, this distance decrease is not fully in line with the regression coefficient of the $\log d$ term, because here, we do not control for hierarchical patterns.

\begin{figure}[h!]
    \centering
    \includegraphics[width=0.7\textwidth]{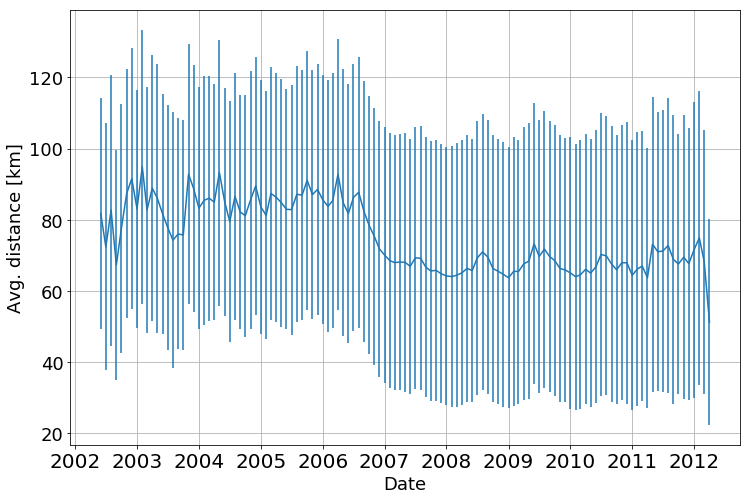}
    \caption{Average distance of invitations in time.}
    \label{fig:avg_dist}
\end{figure}

\clearpage

\noindent The average population size difference between source and target settlement decreases, that might be connected to the decreasing role of hierarchy in the invitation process.

\begin{figure}[h!]
    \centering
    \includegraphics[width=0.7\textwidth]{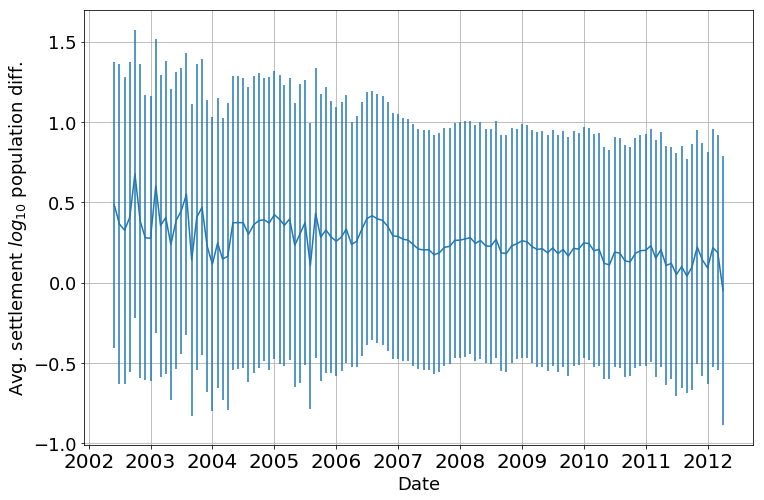}
    \caption{Average population difference of invitations in time.}
    \label{fig:avg_hier}
\end{figure}

\clearpage

\noindent High share of short-range invitations in the middle of the life cycle is in line with the regression coefficient of the distance term from the negative binomial part of the ZINB regression.

\begin{figure}[h!]
    \centering
    \includegraphics[width=0.7\textwidth]{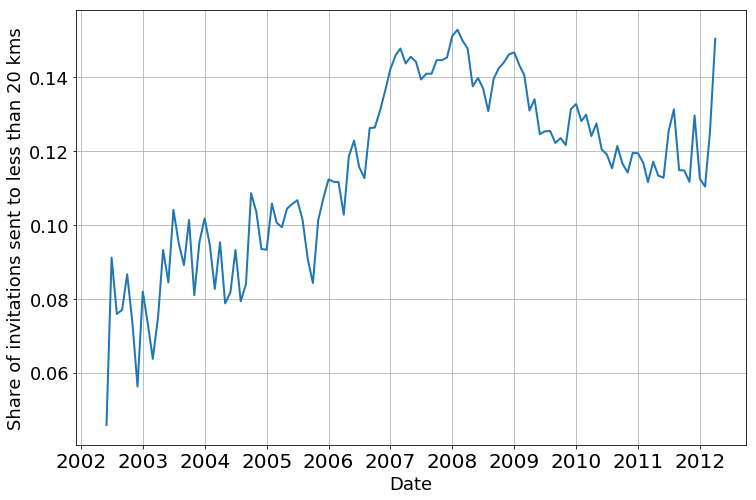}
    \caption{Share of invitations sent to less than 20~km.}
    \label{fig:avg_20km}
\end{figure}

\clearpage

\noindent The share of invitations sent to a settlement at least 3x smaller than the source settlement is constantly decreasing.  In the early and late phases, share of invitations sent into the different direction, to an at least 3x larger settlement than the source settlement is somewhat larger than in the middle of the time range.

\begin{figure}[h!]
    \centering
    \includegraphics[width=0.7\textwidth]{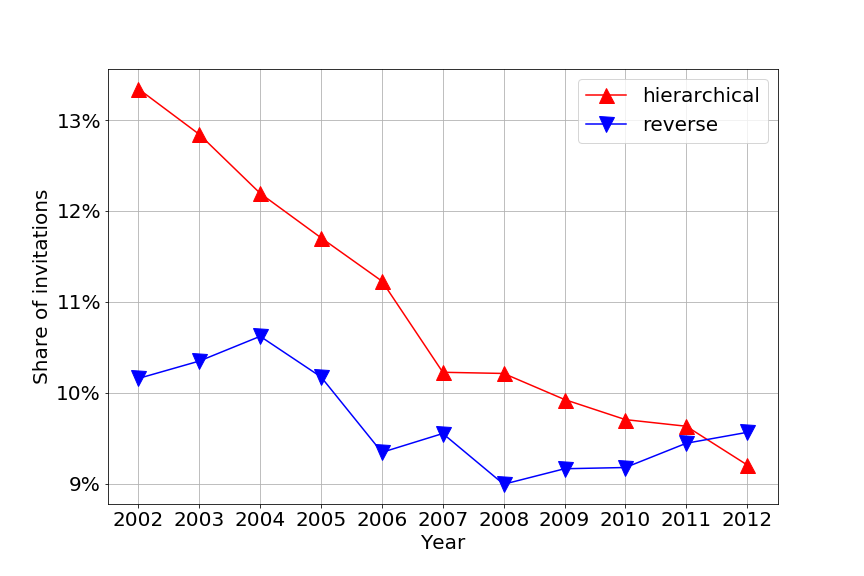}
    \caption{Share of invitations downwards and upwards the hierarchy level (at least ~3x population difference in either direction).}
    \label{fig:hierarchy_agg_years}
\end{figure}

\clearpage

\noindent The yearly distance distributions show a very pronounced tail for the early years indicating the large weight of long-range invitations in the early stages. High share of short-range invitations indicate the presence of neighborhood diffusion.

\begin{figure}[h!]
    \centering
    \includegraphics[width=0.7\textwidth]{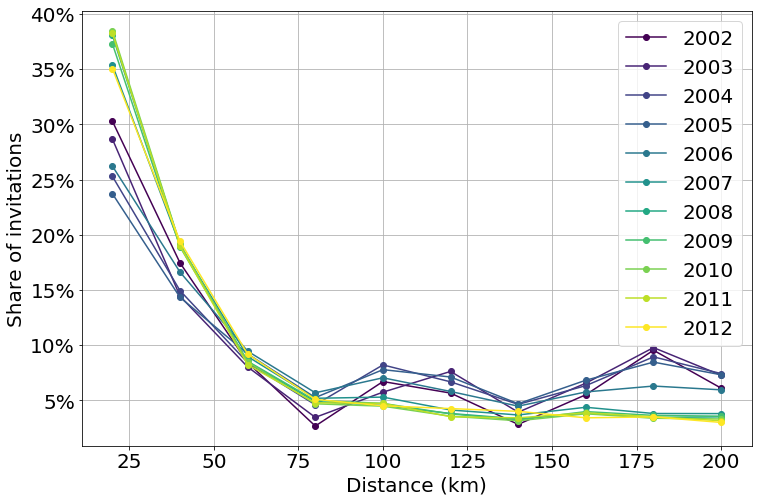}
    \caption{Share of invitations in given distance ranges per year.}
    \label{fig:distance_years}
\end{figure}

\clearpage

\noindent The distribution of hierarchy differences in invitations shows a tendency towards hierarchical diffusion, since there are always more invitations sent to smaller settlements (x axis larger than 0), than invitation sent from smaller to larger settlements (x axis smaller than 0). In the early stages, most invitations go between settlements of very different sizes, then similar hierarchy levels tend to dominate the diffusion process.

\begin{figure}[h!]
    \centering
    \includegraphics[width=0.7\textwidth]{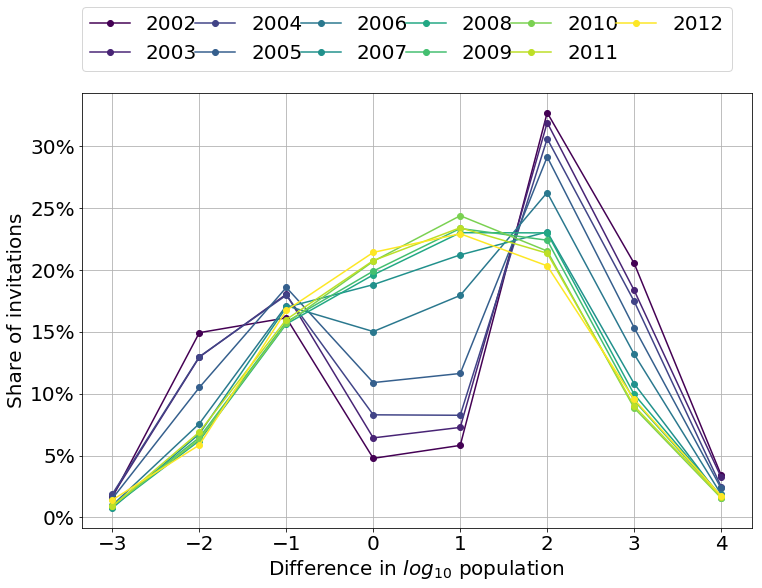}
    \caption{Share of invitations with given $log_{10}$ population difference ranges per year.}
    \label{fig:hierarchy_years}
\end{figure}

\clearpage

\section*{Supplementary Information 3: Regression coefficients}

In the following two tables, we can find the regression coefficients of the full zero-inflated negative binomial model for all years. Significance levels are marked by stars, standard errors of coefficients are given in brackets. The inflation part corresponds to the estimation of the logit probability term $l$ from equation \eqref{eq:fullmodel}, and the negative binomial part to the estimation of the count process term $k$ from \eqref{eq:fullmodel}. Both terms are estimated by the form
\[\alpha (\log P_S)^2 + \beta \cdot (\log P_S \cdot \log P_T) + \gamma (\log P_T)^2 + \delta \log P_S +\varepsilon \log P_T + \chi \log d_{ST} + C,\]
where $\alpha$, $\beta$, $\gamma$, $\delta$, $\varepsilon$, and $\chi$ are the coefficients of the variables listed in the regression tables. Note the significance of the second-order terms in the negative binomial part of the model.

\clearpage

\begin{landscape}

\vfill

\begin{table}[p]
    \footnotesize
    \centering
    \input{regressions/regression_coeffs1}
    \caption{Results of the zero-inflated negative binomial regression models for each year (2002-2007).}
    \label{tab:reg1}
\end{table}

\vfill

\clearpage

\begin{table}[p]
    \footnotesize
    \centering
    \input{regressions/regression_coeffs2}
    \caption{Results of the zero-inflated negative binomial regression models for each year (2007-2012).}
    \label{tab:reg2}
\end{table}

\end{landscape}

\end{document}

%% file: regressions/regression_coeffs1.tex
\begin{tabular}{ll|llllllllllll}
           & Year & \multicolumn{2}{c}{2002} & \multicolumn{2}{c}{2003} & \multicolumn{2}{c}{2004} & \multicolumn{2}{c}{2005} & \multicolumn{2}{c}{2006} & \multicolumn{2}{c}{2007} \\
           &  &       Coeff. &       SE &       Coeff. &       SE &       Coeff. &        SE &       Coeff. &        SE &        Coeff. &        SE &        Coeff. &        SE \\
Model part & Variable &            &           &            &           &            &            &            &            &             &            &             &            \\
\hline \hline
 & Num. obs. &    6525466 &           &    6525466 &           &    6525466 &            &    6525466 &            &     6525466 &            &     6525466 &            \\
           & $\log \alpha$ &  -0.568*** &   (0.176) &  -0.832*** &   (0.147) &  -0.895*** &    (0.126) &  -0.792*** &   (0.0511) &   -0.274*** &   (0.0156) &  -0.0661*** &   (0.0121) \\ \hline
Inflation & Constant &   28.52*** &   (10.84) &   32.24*** &   (5.794) &   23.99*** &    (3.397) &   13.05*** &    (1.433) &   -7.983*** &     (0.78) &   -10.57*** &    (0.665) \\
           & $\log P_S$ &     -2.452 &   (1.514) &  -2.531*** &   (0.629) &  -2.012*** &    (0.339) &   -0.65*** &    (0.178) &    1.298*** &    (0.103) &    1.635*** &   (0.0974) \\
           & $\log P_S^2$ &    0.00491 &   (0.102) &     0.0241 &  (0.0254) &     0.023* &   (0.0119) &  -0.026*** &  (0.00704) &   -0.111*** &  (0.00457) &   -0.129*** &  (0.00512) \\
           & $\log P_S\cdot \log P_T$ &     0.0686 &  (0.0473) &  0.0709*** &   (0.023) &   0.039*** &   (0.0151) &    0.00279 &  (0.00816) &  -0.0171*** &  (0.00622) &  -0.0309*** &  (0.00636) \\
           & $\log P_T$ &   -1.777** &   (0.845) &  -2.201*** &    (0.55) &  -1.215*** &    (0.369) &   -0.69*** &    (0.157) &    1.515*** &    (0.107) &     1.62*** &   (0.0945) \\
           & $\log P_T^2$ &     0.0417 &  (0.0296) &  0.0434*** &  (0.0165) &    0.00774 &   (0.0115) &   -0.00393 &  (0.00505) &   -0.119*** &  (0.00513) &   -0.114*** &   (0.0045) \\
           & $\log d_{ST}$ &   0.383*** &   (0.122) &   0.487*** &   (0.071) &   0.625*** &   (0.0511) &    0.68*** &   (0.0236) &      0.6*** &   (0.0124) &     0.63*** &   (0.0105) \\ \hline
Neg. binom. & Constant &      15.55 &   (10.94) &   22.22*** &   (5.611) &   17.81*** &    (3.306) &   9.638*** &    (1.201) &   -1.285*** &    (0.458) &    -2.62*** &    (0.295) \\
           & $\log P_S$ &  -4.256*** &   (0.982) &  -4.205*** &    (0.47) &  -3.575*** &    (0.301) &  -2.178*** &    (0.122) &   -0.532*** &   (0.0514) &    -0.29*** &   (0.0355) \\
           & $\log P_S^2$ &    0.21*** &  (0.0231) &   0.176*** &  (0.0114) &   0.145*** &  (0.00856) &  0.0955*** &  (0.00358) &   0.0506*** &   (0.0016) &   0.0459*** &  (0.00122) \\
           & $\log P_S\cdot \log P_T$ &       0.06 &  (0.0414) &   0.115*** &  (0.0218) &   0.113*** &   (0.0128) &   0.104*** &  (0.00536) &   0.0625*** &  (0.00252) &   0.0416*** &  (0.00186) \\
           & $\log P_T$ &     -0.778 &   (0.897) &  -1.602*** &   (0.519) &  -1.143*** &    (0.332) &  -0.882*** &    (0.125) &     -0.0645 &   (0.0511) &     0.31*** &   (0.0354) \\
           & $\log P_T^2$ &  0.0649*** &  (0.0187) &  0.0661*** &  (0.0124) &  0.0416*** &  (0.00933) &  0.0369*** &  (0.00358) &   0.0231*** &  (0.00158) &   0.0123*** &   (0.0012) \\
           & $\log d_{ST}$ &  -1.051*** &  (0.0526) &  -0.882*** &   (0.039) &  -0.787*** &   (0.0342) &  -0.918*** &   (0.0161) &   -1.389*** &  (0.00789) &   -1.605*** &  (0.00659) \\ \hline
           \multicolumn{14}{r}{(* = $p<0.10$, **=$p<0.05$,  ***=$p<0.01$)}
\end{tabular}

%% file: regressions/regression_coeffs2.tex
\begin{tabular}{ll|llllllllllll}
           & Year & \multicolumn{2}{c}{2008} & \multicolumn{2}{c}{2009} & \multicolumn{2}{c}{2010}  & \multicolumn{2}{c}{2011} & \multicolumn{2}{c}{2012} \\
           &  &       Coeff. &       SE &       Coeff. &       SE &       Coeff. &        SE &       Coeff. &        SE &        Coeff. &        SE \\
Model part & Variable &            &           &            &           &            &            &            &            &             &            &             &            \\
\hline \hline
 & Num. obs. &    6525466 &           &    6525466 &            &    6525466 &            &     6525466 &            &     6525466 &            \\
           & $\log \alpha$ &   -0.123*** &   (0.0171) &   -0.249*** &   (0.0218) &   -0.315*** &   (0.0248) &    -1.24*** &   (0.0788) &     -2.817* &   (1.494) \\ \hline
Inflation & Constant &   -7.207*** &    (0.619) &   -6.497*** &    (0.644) &   -6.925*** &    (0.671) &       0.316 &    (0.952) &    13.23*** &   (3.106) \\
           & $\log P_S$ &    1.212*** &   (0.0869) &    1.237*** &   (0.0902) &    1.153*** &   (0.0913) &    0.379*** &    (0.122) &   -1.171*** &   (0.375) \\
           & $\log P_S^2$ &   -0.094*** &  (0.00409) &   -0.106*** &  (0.00424) &     -0.1*** &  (0.00419) &  -0.0939*** &   (0.0054) &  -0.0482*** &  (0.0173) \\
           & $\log P_S\cdot \log P_T$ &  -0.0207*** &  (0.00577) &      0.0077 &  (0.00578) &    0.0124** &  (0.00565) &   0.0812*** &  (0.00666) &    0.132*** &  (0.0189) \\
           & $\log P_T$ &    1.006*** &   (0.0878) &    0.791*** &   (0.0891) &    0.893*** &   (0.0901) &       0.21* &     (0.12) &      -0.439 &   (0.388) \\
           & $\log P_T^2$ &  -0.0809*** &  (0.00413) &  -0.0821*** &  (0.00418) &  -0.0864*** &  (0.00413) &  -0.0858*** &  (0.00531) &  -0.0886*** &  (0.0174) \\
           & $\log d_{ST}$ &    0.657*** &   (0.0123) &    0.723*** &   (0.0139) &    0.739*** &   (0.0155) &    1.087*** &   (0.0249) &    1.208*** &  (0.0791) \\ \hline
Neg. binom. & Constant &   -1.812*** &     (0.31) &   -1.763*** &    (0.349) &   -2.677*** &    (0.369) &      -1.183 &    (0.725) &    9.663*** &   (3.017) \\
           & $\log P_S$ &     0.00134 &   (0.0374) &     0.0804* &   (0.0416) &    0.213*** &   (0.0438) &      0.0565 &   (0.0809) &   -1.367*** &   (0.318) \\
           & $\log P_S^2$ &    0.031*** &  (0.00134) &   0.0222*** &   (0.0015) &   0.0145*** &  (0.00158) &  -0.00572** &  (0.00285) &     0.0211* &  (0.0116) \\
           & $\log P_S\cdot \log P_T$ &    0.034*** &  (0.00194) &   0.0386*** &   (0.0021) &   0.0362*** &  (0.00219) &   0.0748*** &  (0.00358) &    0.141*** &  (0.0114) \\
           & $\log P_T$ &    -0.096** &   (0.0387) &   -0.223*** &   (0.0428) &   -0.144*** &   (0.0445) &   -0.442*** &   (0.0805) &   -1.447*** &   (0.322) \\
           & $\log P_T^2$ &   0.0301*** &  (0.00139) &   0.0318*** &  (0.00154) &   0.0275*** &   (0.0016) &   0.0143*** &  (0.00279) &     0.0218* &  (0.0117) \\
           & $\log d_{ST}$ &   -1.566*** &  (0.00909) &   -1.476*** &   (0.0111) &   -1.442*** &   (0.0131) &   -1.017*** &   (0.0212) &   -0.709*** &   (0.066) \\ \hline
           \multicolumn{14}{r}{(* = $p<0.10$, **=$p<0.05$,  ***=$p<0.01$)}
\end{tabular}